\documentclass[intlimits,twoside,a4paper]{article}
\usepackage{amsmath,amssymb}
\usepackage{graphicx}
\usepackage{wrapfig}

\usepackage[T2A]{fontenc}
\usepackage[cp1251]{inputenc}

\usepackage[eqsecnum]{cmpj2}

\issue{2012}{15}{4}{43703}

\doinumber{10.5488/CMP.15.43703}

\title[Spin torque AFM nanooscillator]%
{Spin torque antiferromagnetic nanooscillator \\ in the presence of magnetic noise}

\author[H.~Gomonay, V.~Loktev]{H.~Gomonay\refaddr{label1,label2},
        V.~Loktev\refaddr{label2}}
\addresses{
\addr{label1} National Technical University of Ukraine ``Kiev Polytechnic Institute'',
37 ave Peremogy, 03056 Kyiv, Ukraine %
\addr{label2} Bogolyubov Institute for Theoretical Physics
 of the National Academy of Sciences of Ukraine, \\ 14--b Metrologichna St., 03680 Kyiv, Ukraine}

\date{Received July 3, 2012, in final form August 22, 2012}

\authorcopyright{H.~Gomonay, V.~Loktev, 2012}

\begin{document}

\maketitle
\begin{abstract}
Spin-torque effects in antiferromagnetic (AFM) materials are of
great interest due to the possible applications as high-speed
spintronic devices. In the present paper we analyze the
statistical properties of the current-driven AFM nanooscillator
that result from the white Gaussian noise of magnetic nature.
According to the peculiarities of deterministic dynamics, we
derive the Langevin and Fokker-Planck equations in the energy
representation of two normal modes. We find the stationary
distribution function in the subcritical and overcritical regimes
and calculate the current dependence of the average energy, energy
fluctuation and their ratio (quality factor). The noncritical mode
shows the Boltzmann statistics with the current-dependent
effective temperature in the whole range of the current values. The
effective temperature of the other, i.e., soft, mode critically depends
on the current in the subcritical region. Distribution function of the
soft mode follows the Gaussian law above the generation threshold.
In the overcritical regime, the total average energy and the quality
factor grow with the current value. This raises the AFM
nanooscillators to the promising candidates for active
spintronic components.
%%p----------------------------

\pacs 75.50.Ee, 85.75.-d,  05.40.Ca, 05.10.Gg, 72.25.Pn

\end{abstract}

\section{Introduction}
Nowadays the spin-polarized current is widely used for
manipulation of nano-magnetic structures. Corresponding physical
mechanism is based on the spin-transfer-torque (STT) effect
predicted by Slonczewski and Berger~\cite{Slonczewski:1996,
Berger:1996}: a spin-polarized current may transfer an angular
momentum to a free ferromagnetic (FM) layer and  produce a
macroscopic torque on the latter’s magnetization. In the small
magnetically uniform FM particles, the spin transfer torque induces
a steady rotation of magnetization (see, e.g.
\cite{Kiselev:2003, Tulapurkar:2005, Slavin:2007PhRvB..76b4437G,
Yamaguchi:2008PhRvB..78j4401Y}). Interesting applications of this
effect include: spintronic diodes competing with electronic ones,
radio-frequency devices used for telecommunications, timing
mechanisms.

Recently it was shown~\cite{gomo:2010} (see also
\cite{Linder:2011PhRvB..84i4404L}) that the STT effect should also
occur in antiferromagnetic (AFM) materials  and, by analogy with
FMs, should induce steady rotation of the N\'eel (or AFM) vector.
Current-controlled AFM nanoparticles are the promising candidates
for spintronic devices due to high working frequencies that
fall into $0.1\div1$~THz range (for comparison, typical frequencies of
FM nanooscillators are $1\div50$~GHz~\cite{Rippard:2004, Deac:2008}).
The practical applications, however, face the challenge to improve
the quality factor of nanooscillators and to reduce and control
their linewidth. Thus, to handle and operate  STT devices, we
need to understand the stochastic processes (such as thermal
noise) that set conditions for a linewidth.

In his seminal paper~\cite{Brown:PhysRev.130.1677} W.F.~Brown
first has described the thermal noise in FM particles as a
stochastic magnetic field acting on the magnetization and
has derived the corresponding Fokker-Planck equation which was later
generalized by
 Apalkov and
Visscher~\cite{Apalkov:PhysRevB.72.180405} to the case of
Slonczewski STT. Since then, the noise properties of the FM-based
nonlinear oscillator in the presence of spin-polarized current
were studied both experimentally and theoretically
\cite{Slavin:2007ApPhL..91s2506T, Chudnovskiy:PhysRevB.82.144404,
Dunn:2011ApPhL..98n3109D, Prokopenko2011ApPhL..99c2507P}. Some
authors~\cite{Raikher:2008E, Ouari:PhysRevB.83.064406,
Mishra:springerlink:10.1140/epjb/e2010-00293-0,
Mishra2011PhRvB..84b4429M} used the Brown's approach
 to describe superparamagnetism and switching processes in AFM nanoparticles. Corresponding models, however, considered the particles as weak
ferromagnets and rested on the FM
  moment that inevitably
arises from the surface effects or Dzyaloshinskii-Moriya
interactions.

Analysis of noise in AFM systems requires more complicated
formalism as compared with FMs due to: (i) a larger
number degrees of freedom; (ii) Newtonian-like (vs
precession-like in FMs) dynamics of the N\'eel vector. Moreover,
the peculiarities of AFM dynamics related to a  strong exchange
coupling between the magnetic sublattices, --- magnetoelastic
effects~\cite{Borovik-Romanov:1984}, spin-wave spectra~\cite{Bar:1968E}, STT phenomena~\cite{gomo:2010}, --- should be
described in terms of the variables inherent to AFM ordering.

 In the present paper we investigate
the efficiency of the current-driven AFM nanooscillator in the
framework of the approach based on the N\'eel vector dynamics~\cite{gomonay:2012arXiv1207.1997G}. We assume that the thermal
noise in AFM particle arises from fluctuations of the random
magnetic field. Following the method of slow and fast variables~\cite{Chudnovskiy2011arXiv1110.3750D} and energy representation
for non-equilibrium steady state~\cite{Lev:2010PhRvE..82c1101L} we
formulate the Fokker-Planck equation for energy distribution and
study the linewidth of spin-torque AFM nanooscillator depending on
temperature and current.
%%------------------

\section{Magnetic dynamics of antiferromagnetic particle in the presence of spin-polarized current}
\begin{figure}[!b]
\centerline{
\includegraphics[width=0.65\textwidth]{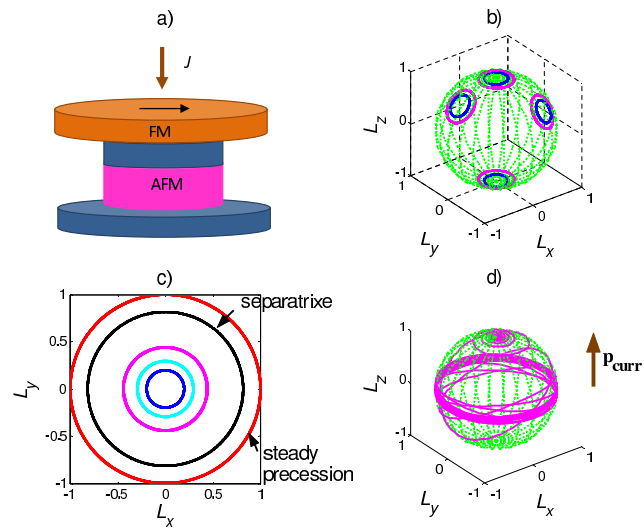}    %{fig_1_lviv_v3}
}
\caption{(Color online) The dynamics of AFM vector in the presence of spin-polarized
current. a) AFM particle is placed between two electrodes, the top
electrode being FM (polarizer). The spacer between FM and AFM is
nonmagnetic to avoid the direct exchange coupling between the
magnetic layers. b) Typical trajectories of $\mathbf{L}$ vector
for circularly polarized modes. Different areas correspond to
different equilibrium orientations (4 of 6 possible are shown).
c) Projection of trajectories in $\mathbf{L}$ space to the $xy$
plane. Three inner circles correspond to the normal modes with
different amplitudes in the vicinity of equilibrium state
$\mathbf{L}\|z$. The last but one circle corresponds to separatrix
and the outer circle is a steady state trajectory. d) Trajectory
in the overcritical regime ($J=1.2J_{\mathrm{cr}}$). In the
initial state $\mathbf{L}\|\mathbf{p}_{\mathrm{curr}}\|z$.}
\label{fig_1_lviv}
\end{figure}

We study an AFM nanoparticle [see figure~\ref{fig_1_lviv}~(a)] large
enough to ensure the AFM ordering and small enough  to neglect the
space variation of the magnetic properties (macrospin
approximation).  The ancillary elements, hard FM (polarizer) and
thin nonmagnetic layers deliver the spin-polarized current to
AFM. We assume that the temperature $T$ is constant and neglect
the thermal (Joule) heating of the system.

We consider a collinear AFM with two equivalent magnetic
sublattices $\mathbf{M_1}$ and $\mathbf{M}_2$ and disregard a weak
FM moment that can arise either from the intrinsic properties of
material or from the surface effects. The coupling between the
sublattices (characterized with a spin-flip field  $H_E$) is
 much stronger than the external fields and the magnetic
anisotropy. To this end, the AFM vector,
$\mathbf{L}\equiv\mathbf{M_1}-\mathbf{M}_2$ of a fixed length
$|\mathbf{L}|=2M_0$ unanimously describes the magnetic state of
AFM nanoparticle. Corresponding dynamic equations for $\mathbf{L}$
are obtained within the standard Lagrange technique with the
Lagrange function~\cite{Bar-june:1979E}
\begin{equation} \label{Lagrangian_AFM} \mathcal{L}_{\mathrm{AFM}}
= \frac{m_L}{2}\dot{\mathbf{L}}^2+\gamma
m_L\left[\dot{\mathbf{L}}\cdot(\mathbf{L}\times\mathbf{H})\right]-
w_{\mathrm{an}}(\mathbf{L})+\frac{\gamma^2m_L}{2}(\mathbf{L}\times{\mathbf{H}})^2,
\end{equation}
where $\mathbf{H}$ is an external magnetic field and
$w_{\mathrm{an}}(\mathbf{L})$ is the  energy of magnetic
anisotropy that forms a potential well for AFM vector, $\gamma$ is
the gyromagnetic ratio, $m_L$ is an effective mass related with
the magnetic susceptibility.

The AFM layer has a cubic symmetry and magnetic anisotropy is
modeled as follows:
\begin{equation}\label{anisotropy}
w_{\rm an}=-\frac{H_{\mathrm{an}}}{8M_0^3}\left(L_x^4+L_y^4+L_z^4\right)\,,
\end{equation}
where the orthogonal axes $x$, $y$ and $z$ coincide with the easy
directions for the N\'eel vector, $H_{\mathrm{an}}\ll H_E$ is
anisotropy field.

The
 Rayleigh dissipation function in the presence of spin-polarized current $J$ takes the form
\cite{gomo:2010}:
\begin{equation}
\label{Relay} \mathcal{R}_{\mathrm{AFM}} =
\gamma_{\mathrm{AFM}}m_L\dot{\mathbf{L}}^2-\frac{\sigma J}{2\gamma
M_0}\left[\mathbf{p}_{\mathrm
{curr}}\cdot\left(\mathbf{L}\times\dot{\mathbf{L}}\right)\right].
\end{equation}
Here, the first term models the internal damping with the
coefficient $2\gamma_{\mathrm{AFM}}$ equal to AFMR linewidth, the
constant $\sigma =\hbar
\gamma\varepsilon/(2eM_{0}v_{\mathrm{AFM}})$ is proportional to
the efficiency $\varepsilon$ of the spin transfer processes,
$v_{\mathrm{AFM}}$ is the volume of AFM nanoparticle, $\hbar$ is
the Plank constant, $e$ is the electron charge. Unit vector
$\mathbf{p}_{\mathrm {curr}}$ is parallel to the spin current
polarization.

Deterministic current-induced dynamics of AFM was analyzed in
detail in reference~\cite{gomo:2010}. Here we reproduce some
characteristic features of this behaviour assuming that
polarization of the spin current is parallel to an easy axis,
$\mathbf{p}_{\mathrm {curr}}\|z$.

First, the critical current
$J_{\mathrm{cr}}\equiv2\gamma_{\mathrm{AFM}}\Omega_{\mathrm{AFMR}}/(\gamma\sigma
H_E)$ that separates the equilibrium and stationary regimes depends
upon the AFMR frequency $\Omega_{\mathrm{AFMR}}\equiv
\gamma\sqrt{H_EH_{\mathrm{an}}}$.

Second, in the subcritical regime, $|J|<J_{\mathrm{cr}}$, the AFM
vector has three equilibrium orientations slightly deflected (due
to STT) from $x$, $y$ and $z$ axes that define six (corresponding
to $\mathbf{L}$ and $-\mathbf{L}$) basins of finite motion in the
phase space. Typical phase trajectories in the vicinity of
equilibrium points in nondissipative approximation correspond to
the clockwise/counterclockwise rotations of AFM vector [circles in
figure~\ref{fig_1_lviv}~(b)] and could be associated with the
circular polarized normal modes with the frequencies
$\pm\Omega_{\mathrm{AFMR}}$. Spin-polarized current acts as a
negative damping for one of the modes (``soft'' mode) and as a
positive damping for the other. The internal losses, however,
suppress the negative damping and the real phase trajectories are
the twisted spirals.

Third, in the overcritical regime ($|J|>J_{\mathrm{cr}}$), the
stable non-equilibrium state occurs when the current-induced
energy pumping exactly compensates the internal losses. AFM vector
rotates in $xy$ plane (perpendicular to $\mathbf{p}_{\mathrm
{curr}}$) with the current-dependent frequency
$\omega=(J/J_{\mathrm{cr}})\Omega_{\mathrm{AFMR}}$ [the outer
circle in fi\-gu\-re~\ref{fig_1_lviv}~(c)]. The rotation direction
coincides with that of the soft mode.
 This regime can be associated with the power
generation.

Fourth, the motion of AFM vector shows two well separated time scales:
fast rotation around $z$ axis with the frequency $\propto
\Omega_{\mathrm{AFMR}}$ and slow, with the characteristic time
$\propto 1/\gamma_{\mathrm{AFM}}$ variation of the polar angle
from $0$ ($\mathbf{L}\|z$) to $\pi/2$ ($\mathbf{L}\perp z$).
Figure~\ref{fig_1_lviv}~(d) shows an example of a typical trajectory
with different time scales for the current-induced transition from
the equilibrium state $\mathbf{L}\|z$ to the steady precession in
$xy$ plane. Deviation from this scenario takes place only in the
close vicinity of the separatrix [see figure~\ref{fig_1_lviv}~(c),
next to the last circle] where the rotation frequency substantially
diminishes.

Thus, in the vicinity of equilibrium and stationary steady states,
the current-induced dynamics of AFM is characterized with a set
of slow and fast variables and this can help to simplify the
description of the stochastic behaviour.

\section{Langevin dynamics and Fokker-Planck equations in energy representation} %???
While only two variables in configuration space are enough to
describe the dynamics of FM nanoparticle, the phase space of AFM
nanoparticle includes  at least four variables: generalized coordinates
$\mathbf{L}$ and corresponding generalized momenta $\mathbf{P}_L$
(with account of normalization condition
$|\mathbf{M}_1|=|\mathbf{M}_2|=M_0$ far below the N\'eel
temperature). This results in rather complicated Langevin
equations~\cite{gomonay:2012arXiv1207.1997G}:
\begin{eqnarray}\label{eq_stochastic_Langevin}
  \dot{\mathbf{L}} &=& \mathbf{P}_L/m_L -\gamma\mathbf{L}\times\mathbf{h}\,,\nonumber\\
  \dot{\mathbf{P}}_L &=& \mathbf{F}_L+\mathbf{F}_{\mathrm{diss}}
  -\gamma\left(\mathbf{P}_L-2\gamma_{\mathrm{AFM}}m_L\mathbf{L}\right)\times\mathbf{h}\,,
\end{eqnarray}
where $\mathbf{F}_L\equiv-\partial
w_{\mathrm{an}}(\mathbf{L})/\partial \mathbf{L}$ is the potential
(gradient) force, and the dissipative force
$\mathbf{F}_{\mathrm{diss}}$ is expressed as follows:
\begin{equation}\label{eq_dissipative_force}
    \mathbf{F}_{\mathrm{diss}}\equiv-\left.\frac{\partial \mathcal{R}_{\rm AFM}}{\partial\dot{\mathbf{L}}}\right|_{\dot{\mathbf{L}}\rightarrow \mathbf{P}_L}=-2\gamma_{\mathrm{AFM}}\mathbf{P}_L-\frac{\sigma
J}{2\gamma M_0}\mathbf{p}_{\mathrm {curr}}\times\mathbf{L}\,.
\end{equation}

The random magnetic field $\mathbf{h}(t)$ in equation
(\ref{eq_stochastic_Langevin}) models the white Gaussian noise
with
\begin{equation}\label{eq_noise_model}
    \langle \mathbf{h}(t)\rangle=0,  \qquad  \langle h_j(t_1)h_k(t_2)\rangle=2D\delta_{jk}\delta(t_1-t_2),
\end{equation}
where $2D$ represents the intensity of thermal fluctuations.

\begin{figure}[!ht]
\centerline{
\includegraphics[width=0.65\textwidth]{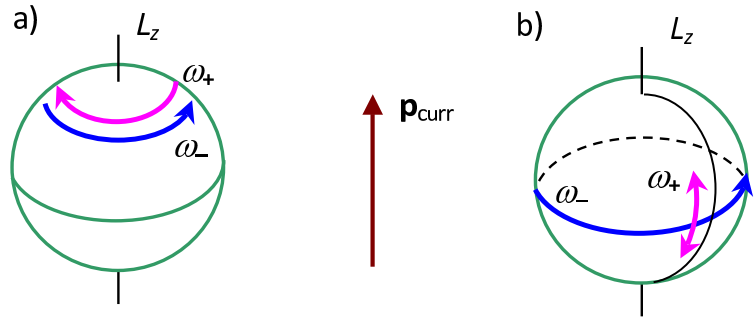}%{fig_4_lviv}
}
\caption{(Color online) Normal circularly polarized modes in the presence of
spin-polarized current in a) subcritical and b) overcritical
regimes. Typical trajectories lie over the sphere
$|\mathbf{L}|=2M_0$ in configuration space.} \label{fig_4_lviv}
\end{figure}

The above mentioned peculiarities of the current-induced dynamics
make it possible  to reduce the number of the effective phase variables in
Langevin, (\ref{eq_stochastic_Langevin}), and corresponding
Fokker-Planck equations. First, in the vicinity of equilibrium and
stationary states, the motion of AFM vector is finite and can be
decomposed to a linear combination of two independent (normal)
modes shown in figure~\ref{fig_4_lviv}. Thus, one can use a set
of canonically conjugated variables ``action-angle'', $I_\pm,
\varphi_\pm$ ($\pm$ correspond to the clock/coun\-ter\-clock\-wise
rotations in configuration space), instead of coordinates and
momenta. Second, in a low-temperature approximation, when the
temperature $T$ is much less than the energy barrier between the
different equilibrium/stationary states, all the essential phase
trajectories for each mode are degenerated (have the same
rotation/oscillation frequency, $\omega_\pm$). Thus, instead of
action, one can use the energy of the mode,
$E_\pm=I_\pm\omega_\pm$, as a canonical variable. Third, two time
scales allow one to exclude the angle variables $\varphi_\pm$ by
averaging over the period of rotation.

As a result, Langevin equations (\ref{eq_stochastic_Langevin})
take the form:
\begin{equation}\label{eq_energy_Langevin}
  \frac{\rd E_{\pm}}{\rd t}=-\overline{\mathbf{P}_L\cdot\mathbf{F}_{\mathrm{diss}}}+\gamma\left[\left(2\gamma_{\mathrm{AFM}}\mathbf{P}_L-\frac{\partial
w_{\mathrm{an}}}{\partial \mathbf{L}
}\right)\cdot\mathbf{L}\times\mathbf{h}\right],
\end{equation}
where the overline means average over the period of rotation and
the summands in r.h.s. of equation~(\ref{eq_energy_Langevin})
should be expressed in terms of the average energies
\begin{equation}\label{eq_average_energy}
  E_{\pm}=\frac{\omega_\pm}{2\pi}\int_{0}^{2\pi/\omega_\pm}\left[\mathbf{P}_L^2/(2m_L)+w_{\mathrm{an}}(\mathbf{L})\right]\rd t.
\end{equation}
An explicit closed form of the equation (\ref{eq_average_energy})
will be obtained in next subsections in the limiting  subcritical
and overcritical regimes.

\subsection{Subcritical regime $J<J_{\mathrm{cr}}$}
Let us consider the basin of states in the vicinity of equilibrium
point $\mathbf{L}\|\mathbf{p}_{\mathrm{curr}}\|z$ and choose $L_x,
L_y\ll 2M_0$ as generalized coordinates, $L_z\approx 2M_0$. Time
dependence of the dynamic variables for the normal modes is given
by the expressions
\begin{equation}\label{eq_normal_modes_subcritical}
 L_x+ \ri L_y=2M_0c_\pm \re^{\ri\omega_\pm t},\qquad
 P_{Lx}+\ri P_{Ly}=2\ri M_0m_L\omega_\pm c_\pm \re^{\ri\omega_\pm
 t},\qquad \omega_\pm=\pm\Omega_{\mathrm{AFMR}}\,.
\end{equation}
The average energy is related with the amplitude $c_\pm$ as
follows: $E_\pm= 2E_0c_\pm^2$, where the value
$E_0\equiv2M^2_0\Omega_{\mathrm{AFMR}}^2m_L=M_0H_{\mathrm{an}}$
defines the characteristic energy scale for AFM nanoparticle.

 Substituting expressions (\ref{eq_normal_modes_subcritical})
 into (\ref{eq_dissipative_force}) and (\ref{eq_energy_Langevin})
 we arrive at the system of two independent Langevin equations:
 \begin{eqnarray}\label{eq-Langevin_amp_subcrit}
  \frac{\rd\varepsilon_{\pm}}{\rd t}&=&{}-2\gamma_{\mathrm{AFM}}\left(1\pm\frac{J}{J_{\mathrm{cr}}}\right)\varepsilon_{\pm}+2\gamma\sqrt{\varepsilon_{\pm}}\left(\pm h_x\sin\Omega_{\mathrm{AFMR}}t-h_y\cos\Omega_{\mathrm{AFMR}}t\right)\\
  &&{}+\gamma\frac{2\gamma_{\mathrm{AFM}}}{\Omega_{\mathrm{AFMR}}}\left[2\sqrt{\varepsilon_{\pm}}\left(h_x\cos\Omega_{\mathrm{AFMR}}t\pm h_y\sin\Omega_{\mathrm{AFMR}}t\right)
  \mp\varepsilon_{\pm}h_z\right],\nonumber
\end{eqnarray}
where $\varepsilon_{\pm}\equiv E_\pm/E_0$ is a dimensionless
energy.

As it is seen from equation (\ref{eq-Langevin_amp_subcrit}), the
clock/counterclockwise modes [figure~\ref{fig_4_lviv}~(a)] interact
with the current in different ways. If $J>0$, the effective
damping of the first mode (with the energy $E_+$) increases and
that of the second (with the energy $E_-$) decreases, due to the
action of spin-polarized current.

Equations (\ref{eq-Langevin_amp_subcrit}) are typical Langevin
equations in energy representation considered in detail in~\cite{Lev:2010PhRvE..82c1101L}. The first summand in the r.h.s.
describes the rate of direct energy exchange which depends
upon the current value $J$. All but first summands in the r.h.s.
of~(\ref{eq-Langevin_amp_subcrit}) account for system-environment
interaction induced by the field fluctuations. The diffusion
functions (coefficients before $h_j$) depend on energy and thus
correspond to multiplicative noise. The last two terms (in square
brackets) are multiplied by a small factor
$\gamma_{\mathrm{AFM}}/\Omega_{\mathrm{AFMR}}\ll 1$ and thus can
be neglected.

In the accepted approximation of noninteracting modes, the
distribution function in phase space,
$f(\mathbf{L},\mathbf{P}_L;t)$, can be factorized as follows:
$f(\mathbf{L},\mathbf{P}_L;t)\Rightarrow f_+(E_+;t)f_-(E_-;t)$.
The Fokker-Planck equations for $f_\pm$  are deduced in a
standard manner from (\ref{eq-Langevin_amp_subcrit}) in
Stratonovich convention and  take the form:
\begin{equation}\label{eq_Fokker-PLank_modes}
  \frac{\partial f_\pm(E_\pm)}{\partial t}=\frac{\partial }{\partial
  E_\pm}\left\{\left[\gamma^2DE_0\sqrt{E_\pm}\frac{\partial }{\partial
  E_\pm}\sqrt{E_\pm}+2\gamma_{\mathrm{AFM}}\left(1\pm\frac{J}{J_{\mathrm{cr}}}\right)E_\pm \right]f_\pm(E_\pm)\right\}.
\end{equation}
From the stationary solution of (\ref{eq_Fokker-PLank_modes}) one
gets the AFM probability distribution function $f(E_+,E_-)=f_+(E_+)f_-(E_-)\sqrt{E_+E_-}$ (with due account of Jacobian):
\begin{equation}\label{eq_distribution_subcritical}
f(E_+,E_-)=f_0\exp{\left\{-\frac{2\gamma_{\mathrm{AFM}}}{\gamma^2DE_0}\left[\left(1+\frac{J}{J_{\mathrm{cr}}}\right)E_++\left(1-\frac{J}{J_{\mathrm{cr}}}\right)E_-\right]\right\}},
\end{equation}
where $f_{0}$ is a normalization constant and the diffusion
coefficient is related with the temperature $T$ through the
fluctuation-dissipation theorem as follows:
\begin{equation}\label{eq_diffusion_coef}
  D=\frac{2\gamma_{\mathrm{AFM}}}{\gamma^2E_0}T.
\end{equation}

\subsection{Overcritical regime $J>J_{\mathrm{cr}}$}
In the overcritical regime, the motion can be decomposed (like it
was done in the reference~\cite{Denisov:2006PhRvB..74j4406D} for
FM) into steady rotation of AFM vector in $xy$ plane with the frequency
$\omega_-$ and small meridian oscillations of $\mathbf{L}$ vector
with the frequency $\omega_+$ [see figure~\ref{fig_4_lviv}~(b)].

The ``$+$''-mode is parametrized as follows:
\begin{eqnarray}\label{eq_normal_modes_supercritical+}
&& L_z \ =c \re^{\ri\omega_+ t}, \qquad\qquad\qquad\quad  P_{Lz}=2\ri M_0m_L\omega_+c \re^{\ri\omega_+
 t},\\
&& \omega_+=\Omega_{\mathrm{AFMR}}\sqrt{\frac{3}{4}+\frac{J^2}{J^2_{\mathrm{cr}}}}, \ \,
\qquad E_+=2E_0\left(\frac{3}{4}+\frac{J^2}{J^2_{\mathrm{cr}}}\right)c^2.\nonumber
\end{eqnarray}
The corresponding Langevin and Fokker-Planck equations are analogous
to (\ref{eq-Langevin_amp_subcrit}) and (\ref{eq_Fokker-PLank_modes}).

Parametrization of ``$-$''-mode coincides with that in equation~(\ref{eq_normal_modes_subcritical}) with $c_-=1$. In this particular case, the proper dynamic variable is action $I_-=8\pi M_0^2m_L\omega_-$. The corresponding Langevin equation takes the form
\begin{equation}\label{eq_Langevin_action_overcritial}
    \frac{\rd I_-}{\rd t}=-2\gamma_{\mathrm{AFM}}\left(I_--8\pi M_0^2m_L\omega_-^{(0)}\right)-\gamma I_-h_z\,,
\end{equation}
where $\omega_-^{(0)}=-(J/J_{\mathrm{cr}})\Omega_{\mathrm{AFMR}}$
is the frequency of steady rotation. Substituting $I_-=4\pi
M_0\sqrt{2m_LE_-}$ into (\ref{eq_Langevin_action_overcritial}) we
get Langevin equation in energy representation:
\begin{equation}\label{eq_Langevin_energy_overcritial}
    \frac{\rd E_-}{\rd t}=-4\gamma_{\mathrm{AFM}}\left(E_--\frac{J}{J_{\mathrm{cr}}}\sqrt{E_0E_-}\right)-2\gamma E_-h_z\,.
\end{equation}
The corresponding Fokker-Planck equation is deduced as follows:
\begin{equation}\label{eq_Fokker-PLank_modes_supercritical}
  \frac{\partial f_-(E_-)}{\partial t}=\frac{\partial }{\partial
  E_-}\left\{\left[4\gamma^2DE_-\frac{\partial }{\partial
  E_-}E_-+4\gamma_{\mathrm{AFM}}\left(E_--\frac{J}{J_{\mathrm{cr}}}\sqrt{E_0E_-}\right) \right]f_-(E_-)\right\}.
\end{equation}

 Ultimately, the stationary AFM distribution function in the overcritial
 regime takes the form:
\begin{equation}\label{eq_distribution_supercritical}
f(E_+,E_-)=f_0\exp{\left\{-\frac{4E_+}{\left[3+4\left(J/J_{\mathrm{cr}}\right)^2\right]T}
-\frac{\left(E_--E_{0}J^2/J^2_{\mathrm{cr}}\right)^2}{2TE_{0}J^2/J^2_{\mathrm{cr}}}\right\}},
\end{equation}
where we have taken into account the relation
(\ref{eq_diffusion_coef}) and assumed that $T\ll E_0$.

\section{Discussion}
In the previous section we derived the stochastic equations for
the current-controlled AFM nanoparticle in the low temperature
approximation, $T\ll E_0$. So, the magnetic anisotropy energy
$E_0$ sets the energy scale of the system and limits the validity
of equations~(\ref{eq_distribution_supercritical}). AFMs with high
N\'eel temperature show a characteristic density of magnetic
anisotropy  $10^3\div 10^4$ J/m$^3$ (see, e.g.~\cite{Takahashi:0022-3727-35-19-307} for Mn$_{82}$Ni$_{18}$). So,
 $E_0\propto 10^{-20}\div10^{-19}$~J for the nanoparticles with the typical size
$50\times50\times5$ nm$^3$. Thus, the proposed model can be
applied up to the room temperature, $T_{\mathrm{RT}}=4\cdot
10^{-21}$~J.

Neglecting the swap processes between different basins, we found
two stationary solutions of Fokker-Planck equations: in the
vicinity of equilibrium, (\ref{eq_distribution_subcritical}), and
nonequilibrium steady (\ref{eq_distribution_supercritical})
states.

For the noncritical (``$+$'') mode, the probability function
$f(E_+)$ [see (\ref{eq_distribution_subcritical}) and
(\ref{eq_distribution_supercritical})] follows the Boltzmann law
with the current-dependent effective temperature
\begin{equation}\label{eq_effective temeprature+}
  T_{\mathrm{eff}}^{(+)}=T\left\{\begin{array}{cc}
    \left(1+ J/J_{\mathrm{cr}}\right)^{-1}\,, & \qquad J<J_{\mathrm{cr}}\,, \\
    0.75+J^2/J^2_{\mathrm{cr}}\,, & \qquad J>J_{\mathrm{cr}}\,.
  \end{array}\right.
\end{equation}

In the subcritical regime ($J<J_{\mathrm{cr}}$), the ``soft''
(``$-$'') mode shows the Boltzmann-like distribution which changes
to the Gauissian-like with the current-dependent average,
$E_-^{(0)}\equiv E_0J^2/J^2_{\mathrm{cr}}$, and current-depen\-dent
dispersion, $TE_0J^2/J^2_{\mathrm{crit}}$ at $J>J_{\mathrm{cr}}$.
In the subcritical region, the effective temperature
$T_{\mathrm{eff}}^{(-)}=T/(1- J/J_{\mathrm{cr}})$ diverges as
$J\rightarrow J_{\mathrm{cr}}-0$. This singularity can be avoided
with due account of the swap processes [see the trajectory in figure~\ref{fig_1_lviv}~(d)] that are important in the vicinity of critical
current $J\approx J_{\mathrm{cr}}$. This problem is, however, out
of scope of this paper.

The statistical properties of the ``soft'' mode in AFM and FM
nanoparticles in the presence of spin-polarized current are
similar: Boltzmann-like distribution and critical behaviour of the
effective temperature in subcritical region, and Gaussian-like
distribution in overcritical regime~\cite{Slavin:2007ApPhL..91s2506T}. The current dependencies of
average energy and dispersion are, however, different, as will be
discussed below.

\begin{figure}[t]
\centerline{
\includegraphics[width=0.5\textwidth]{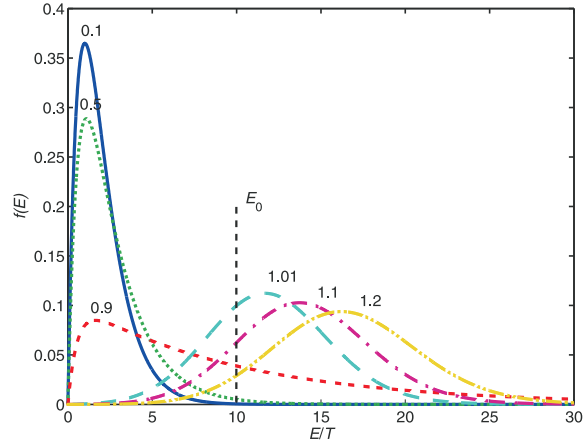}    %{fig_2_lviv_v4}
}
\caption{(Color online) Energy distribution function for different current values
$J/J_{\mathrm{cr}}$ (shown with numbers near the curves) in the
subcritical ($J<J_{\mathrm{cr}}$) and overcritical
($J>J_{\mathrm{cr}}$) regimes. Energy is measured in the
dimensionless units, $E/T$. Vertical dashed line shows the energy
$E_{0}=10T$ in the overcritical regime.} \label{fig_2_lviv}
\end{figure}

Figure~\ref{fig_2_lviv} shows the distribution function, $f(E)$,
that depends on the total magnetic energy $E=E_++E_-$ of AFM
nanoparticle. Dependencies $f(E)$ for different current values are
calculated from (\ref{eq_distribution_subcritical}) and
(\ref{eq_distribution_supercritical}) as conventional
probabilities.

 In the subcritical regime ($J/J_{\mathrm{cr}}=0.1, 0.5,
0.9$), the distribution function is asymmetric with maximum at
\begin{equation}\label{eq_max_subcritical}
  E_{\mathrm{max}}=\frac{T}{J/J_{\mathrm{cr}}}\tanh^{-1}\frac{J}{J_{\mathrm{cr}}}\approx
  T.
\end{equation}

\begin{figure}[t]
\centerline{
\includegraphics[width=0.65\textwidth]{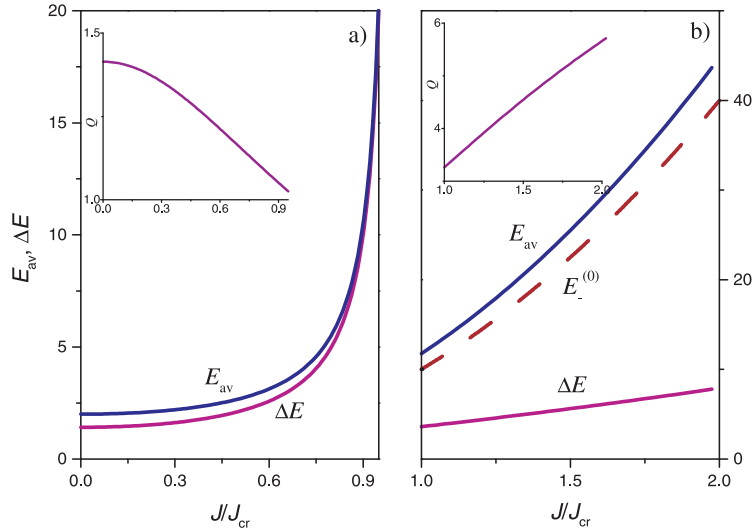}   %{fig_3_lviv_v2}
}
\caption{(Color online) Current dependence of the average energy,
$E_{\mathrm{av}}$, and energy fluctuation, $\Delta E$ in  a)
subcritical ($J<J_{\mathrm{cr}}$) and b)  overcritical
($J>J_{\mathrm{cr}}$) regimes. Insets show the current dependence
of the  quality factor, $Q_{\mathrm{AFM}}=E_{\mathrm{av}}/\Delta
E$. The energy $E_{0}=10T$. Dashed line shows the
current-dependence of the ``soft'' mode energy
$E_-^{(0)}=E_0J^2/J_{\mathrm{cr}}^2$\,.} \label{fig_3_lviv}
\end{figure}

Average magnetic energy of nanoparticle,
$E_{\mathrm{av}}\equiv\langle E\rangle
=T_{\mathrm{eff}}^{(+)}+T_{\mathrm{eff}}^{(-)}$, consists of the
noisy component only and diverges as $J\rightarrow
J_{\mathrm{cr}}$. Energy fluctuation,
\begin{equation}\label{eq_fluctuations}
  \Delta
  E\equiv\sqrt{\langle\left(E-E_{\mathrm{av}}\right)^2\rangle}
  =\left(T_{\mathrm{eff}}^{(+)}+T_{\mathrm{eff}}^{(-)}\right)
  \sqrt{\frac{1}{2}\left(1+\frac{J^2}{J^2_{\mathrm{cr}}}\right)}
\end{equation}
diverges in the same way [see figure~\ref{fig_3_lviv}~(a)]. Thus,
the quality factor, $Q_{\mathrm{AFM}}=E_{\mathrm{av}}/\Delta E$
diminishes down to 1 as $J\rightarrow J_{\mathrm{cr}}-0$. This
tendency is quite obvious if one takes into account the noisy
source of the energy in a system. The singularity in $\Delta  E$
emphasizes the role of thermal noise  in the transition from
equilibrium to nonequilibrium steady state in the vicinity of
critical current.

In the overcritical regime ($J/J_{\mathrm{cr}}=1.01, \, 1.1, \, 1.2$),
the distribution function $f(E)$ has a Gaussian-like shape with
the maximum close to the average energy $E_{\mathrm{av}}$. Both
$E_{\mathrm{av}}$ and $\Delta E$ grow with current [see figure~\ref{fig_3_lviv}~(b)]. However, the main contribution into
$E_{\mathrm{av}}$ arises from the deterministic (low entropy)
current-induced rotation of AFM vector, while $\Delta E$
originates from the noise slightly intensified by the current.
Thus, the quality factor $Q_{\mathrm{AFM}}$ is finite at
$J=J_{\mathrm{cr}}$ and increases with current almost linearly
[inset in figure~\ref{fig_3_lviv}~(b)]. By contrast, the quality
factor for FM nanoparticle,
$Q_{\mathrm{FM}}\propto\sqrt{1-J_{\mathrm{cr}}/J}$, vanishes in
the close vicinity of $J_{\mathrm{cr}}$ in the supercritical
regime. This opens up the way for potential applications of AFM
nanoparticles as active elements of spintronic devices and as a
possible alternative to FM nanooscillators. The model, however,
should be further developed to account for the Joule heating,
current fluctuations, etc.

In summary, we considered the current-induced dynamics of AFM
nanoparticle in the presence of white Gaussian noise which
originates from the random magnetic fields. We found the
stationary energy distribution functions in two regimes:
subcritical, when the spin-polarized current is too small to
reorient AFM vector from the initial equilibrium state, and
overcritical, when the spin-polarized current keeps steady
rotation of AFM vector. Average energy and energy fluctuations in
the subcritical regime show the critical behaviour as
$J\rightarrow J_{\mathrm{cr}}$. This can be used to facilitate
the current-induced reorientation of AFM vector. In the
overcritical regime, the quality factor of AFM particle as
nanooscillator can be increased by adjusting the current value.

\section*{Acknowledgements}
The authors are grateful to B.I. Lev for
fruitful discussions. The paper was partially supported by the
grant from the Ministry of Education, Science, Youth and Sport of
Ukraine and by the Programme of Fundamental Researches of the
Department of Physics and Astronomy of National Academy of
Sciences of Ukraine.

\ukrainianpart

\title{Антиферомагнітний наноосцилятор зі спіновим крутильним моментом в присутності магнітних шумів}
\author{О.В. Гомонай\refaddr{label1,label2}, В.М. Локтєв\refaddr{label2}}
\addresses{
\addr{label1} Національний технічний університет України ``КПІ'', пр. Перемоги, 37, 03056 Київ, Україна
\addr{label2} Інститут теоретичної фізики ім. М.М.~Боголюбова НАН України, \\ вул. Метрологічна,~14-б, 03680 Київ, Україна}
%
%% якщо автор є один або автори є з однієї установи:
%
%  \author{1й Автор, 2й Автор, \ldots}
%  \address{Інститут\ldots}
%
%%

\makeukrtitle

\begin{abstract}
\tolerance=3000%
Процеси передачі крутильного спінового моменту антиферомагнітним (АФМ) матеріалам цікаві з точки зору можливих застосувань в швидкісних спінтронних приладах. В даній роботі вивчаються обумовлені магнітним шумом статистичні властивості АФМ наноосцилятора, який знаходиться під дією спін-поляризованого струму.   Виходячи з особливостей детермінованої динаміки, виведено рівняння Ланжевена та Фоккера-Планка для двох нормальних мод в енергетичному представлені. Магнітний шум моделюється при цьому як випадковий дельта-корельований Гаусів процес. Отримано вирази для стаціонарної функції розподілу в докритичному і надкритичному режимах. Розраховано залежність від струму середньої енергії, флуктуації енергії та їх відношення (фактора якості). Показано, що функція розподілу однієї з мод (некритичної) відповідає Больцманівському розподілу в усьому діапазоні величини струму, причому ефективна температура залежить від струму. Ефективна температура іншої (м'якої) моди залежить від струму критичним чином. В надкритичній області функція розподілу цієї моди відповідає розподілу Гауса, а середня енергія і фактор якості зростають з величиною струму, що робить АФМ наноосцилятори перспективними системами для використання в ролі активних елементів в спінтронних приладах.
\keywords антиферомагнетики, спінтроніка, тепловий шум, рівняння Ланжевена, рівняння Фоккера-Планка, спіновий крутильний момент

\end{abstract}

  \lastpage

\begin{thebibliography}{99}
%\providecommand{\url}[1]{\texttt{#1}}
%\providecommand{\urlprefix}{URL }
%\providecommand{\eprint}[2][]{\url{#2}}

\bibitem{Slonczewski:1996}
Slonczewski J., J. Magn. Magn. Mater., 1996, \textbf{159}, L1; \bibdoi{10.1016/0304-8853(96)00062-5}.

\bibitem{Berger:1996}
Berger L., Phys. Rev. B, 1996, \textbf{54}, No.~13, 9353; \bibdoi{10.1103/PhysRevB.54.9353}.

\bibitem{Kiselev:2003}
Kiselev S.I., Sankey J.C., Krivorotov I.N., Emley N.C., Schoelkopf R.J.,
  Buhrman R.A., Ralph D.C., Nature, 2003, \textbf{425}, 380;   \bibdoi{10.1038/nature01967}.

\bibitem{Tulapurkar:2005}
Tulapurkar A.A., Suzuki Y., Fukushima A., Kubota H., Maehara H., Tsunekawa K.,
  Djayaprawira D.D., Watanabe~N., Yuasa S., Nature, 2005, \textbf{438}, No.~11,
  339; \bibdoi{10.1038/nature04207}.

\bibitem{Slavin:2007PhRvB..76b4437G}
{Gerhart} G., {Bankowski} E., {Melkov} G.A., {Tiberkevich} V.S., {Slavin} A.N.,
  Phys. Rev. B, 2007, \textbf{76}, No.~2, 024437; \bibdoi{10.1103/PhysRevB.76.024437}.

\bibitem{Yamaguchi:2008PhRvB..78j4401Y}
{Yamaguchi} A., {Motoi} K., {Hirohata} A., {Miyajima} H., {Miyashita} Y.,
  {Sanada} Y., Phys. Rev. B, 2008, \textbf{78}, No.~10, 104401; \bibdoi{10.1103/PhysRevB.78.104401}.

\bibitem{gomo:2010}
Gomonay H.V., Loktev V.M., Phys. Rev. B, 2010, \textbf{81}, No.~14, 144427; \bibdoi{10.1103/PhysRevB.81.144427}.

\bibitem{Linder:2011PhRvB..84i4404L}
{Linder} J., Phys. Rev. B, 2011, \textbf{84}, No.~9, 094404; \bibdoi{10.1103/PhysRevB.84.094404}.

\bibitem{Rippard:2004}
Rippard W.H., Pufall M.R., Kaka S., Russek S.E., Silva T.J., Phys. Rev. Lett.,
  2004, \textbf{92}, No.~2, 027201; \\ \bibdoi{10.1103/PhysRevLett.92.027201}.

\bibitem{Deac:2008}
Deac A.M., Fukushima A., Kubota H., Maehara H., Suzukia Y., Yuasa S., Nagamine
  Y., Tsunekawa K., Djayaprawira~D.D., Watanabe N., Nat. Phys., 2008,
  \textbf{4}, 803; \bibdoi{10.1038/nphys1036}.

\bibitem{Brown:PhysRev.130.1677}
Brown W.F., Phys. Rev., 1963, \textbf{130}, 1677;  \bibdoi{10.1103/PhysRev.130.1677}.

\bibitem{Apalkov:PhysRevB.72.180405}
Apalkov D.M., Visscher P.B., Phys. Rev. B, 2005, \textbf{72}, 180405;  \bibdoi{10.1103/PhysRevB.72.180405}.

\bibitem{Slavin:2007ApPhL..91s2506T}
{Tiberkevich} V., {Slavin} A., {Kim} J.V., Appl. Phys. Lett., 2007,
  \textbf{91}, No.~19, 192506; \bibdoi{10.1063/1.2812546}.

\bibitem{Chudnovskiy:PhysRevB.82.144404}
Swiebodzinski J., Chudnovskiy A., Dunn T., Kamenev A., Phys. Rev. B, 2010,
  \textbf{82}, 144404; \\ \bibdoi{10.1103/PhysRevB.82.144404}.

\bibitem{Dunn:2011ApPhL..98n3109D}
{Dunn} T., {Kamenev} A., Appl. Phys. Lett., 2011, \textbf{98}, No.~14, 143109; \bibdoi{10.1063/1.3576929}.

\bibitem{Prokopenko2011ApPhL..99c2507P}
{Prokopenko} O., {Melkov} G., {Bankowski} E., {Meitzler} T., {Tiberkevich} V.,
  {Slavin} A., Appl. Phys. Lett., 2011, \textbf{99}, No.~3, 032507; \bibdoi{10.1063/1.3612917}.

\bibitem{Raikher:2008E}
Raikher Y.L., Stepanov V.I., J. Exp. Theor. Phys., 2008, \textbf{107}, No.~3,
  435; \bibdoi{10.1134/S1063776108090112} [Zh. Eksp. Teor. Fiz., 2008, \textbf{134}, No.~3, 514 (in Russian)].

\bibitem{Ouari:PhysRevB.83.064406}
Ouari B., Kalmykov Y.P., Phys. Rev. B, 2011, \textbf{83}, 064406; \bibdoi{10.1103/PhysRevB.83.064406}.

\bibitem{Mishra:springerlink:10.1140/epjb/e2010-00293-0}
{Mishra S.K., Eur. Phys. J. B, 2010, \textbf{78}, 65; \bibdoi{10.1140/epjb/e2010-00293-0}.}

\bibitem{Mishra2011PhRvB..84b4429M}
{Mishra} S.K., {Subrahmanyam} V., Phys. Rev. B, 2011, \textbf{84}, No.~2,
  024429; \bibdoi{10.1103/PhysRevB.84.024429}.

\bibitem{Borovik-Romanov:1984}
Borovik-Romanov A.S., Rudashevskii E.G., Turov E.A., Shavrov V.G.,
 Sov. Phys. Uspekhi, 1984, \textbf{27}, No.~8, 642; \bibdoi{10.1070/PU1984v027n08ABEH004078}.

\bibitem{Bar:1968E}
Akhiezer A.I., Bar'yakhtar V.G., Peletminskii S.V., Spin Waves, North-Holland Series in Low Temperature Physics, vol.~1, North-Holland, Amsterdam, 1968.

\bibitem{gomonay:2012arXiv1207.1997G}
{Gomonay} H.V., {Loktev} V.M., Eur. Phys. J. ST, 2013 (in press); Preprint \href{http://arxiv.org/abs/1207.1997}{arXiv:1207.1997}, 2012.

\bibitem{Chudnovskiy2011arXiv1110.3750D}
{Dunn} T., {Chudnovskiy} A.L., {Kamenev} A., Preprint \href{http://arxiv.org/abs/1110.3750}{arXiv:1110.3750}, 2011.

\bibitem{Lev:2010PhRvE..82c1101L}
{Lev} B.I., {Kiselev} A.D., Phys. Rev. E, 2010, \textbf{82}, No.~3, 031101; \bibdoi{10.1103/PhysRevE.82.031101}.

\bibitem{Bar-june:1979E}
Bar’yakhtar I., Ivanov B., Fiz. Nizk. Temp., 1979, \textbf{5}, 361 (in Russian).

\bibitem{Denisov:2006PhRvB..74j4406D}
Denisov S.I., Lyutyy T.V., H\"anggi P., Trohidou K.N., Phys. Rev. B, 2006,
  \textbf{74}, 104406; \\ \bibdoi {10.1103/PhysRevB.74.104406}.

\bibitem{Takahashi:0022-3727-35-19-307}
Takahashi M., Tsunoda M., J. Phys. D: Appl. Phys., 2002,
  \textbf{35}, No.~19, 2365;  \doi{10.1088/0022-3727/35/19/307}.

\end{thebibliography}
\end{document}